\documentclass[12pt]{article}

 \newread\testifexists
 \def\GetIfExists #1 {\immediate\openin\testifexists=#1
     \ifeof\testifexists\immediate\closein\testifexists\else
     \immediate\closein\testifexists\input #1\fi}

 \usepackage{gthstyle}\usepackage{amsfonts}
 \usepackage{amssymb}
 \usepackage{graphicx} \usepackage{epstopdf}
 \immediate\openin\testifexists=e
     \ifeof\testifexists\immediate\closein\testifexists\else
     \immediate\closein\testifexists\input e\fipsf

 \def\Bbb#1{\setbox0=\hbox{$\tt #1$}  \copy0\kern-\wd0\kern .1em\copy0}

 \def\bbf#1{\setbox0=\hbox{$#1$} \kern-.025em\copy0\kern-\wd0
         \kern.05em\copy0\kern-\wd0 \kern-.025em\raise.0433em\box0}

 \immediate\openin\testifexists=a
     \ifeof\testifexists\immediate\closein\testifexists\else
     \immediate\closein\testifexists\input a\fimssym.def  

                     \newcommand{\fn}{\footnote}
              
 \newcommand{\be}{\begin{eqnarray}}             \newcommand{\ee}{\end{eqnarray}}
 \newcommand{\bi}[1]{\begin{itemize}\item[#1]}         \newcommand{\itm}[1]{\item[#1]}  \newcommand{\ei}{\end{itemize}}
 \newcommand{\eqn}[1]{(\ref{#1})}

 \newcommand{\crlb}[1]{\label{#1}\\[2pt]}
 \newcommand{\eela}[1]{\,\hbox{\scriptsize{#1}\qquad}\label{#1}\end{eqnarray}}
 \newcommand{\eelb}[1]{\label{#1}\end{eqnarray}}
 
 \newcommand{\newsecb}[2]{\section{#1}\label{#2}\setcounter{equation}{0}}
 	 	
	 \newcommand{\subsecb}[2]{\subsection{#1}\label{#2}}

 \newcommand{\nolabels} {\def\eel{\eelb} \def\crl{\crlb} \def\newsecl{\newsecb}\def\subsecl{\subsecb}\def\bibiteml{\bibitem}\def\citel{\cite}\def\labell{\label}}

\newcommand\publishversion{\nolabels\setlength{\textheight}{9in}\setlength{\oddsidemargin}{0in}
    \setlength{\textwidth}{6.3in}\setlength{\topmargin}{-0.1in}}

 \def\a{\alpha}      \def\b{\beta}   \def\g{\gamma}      
 \def\d{\delta}          \def\e{\varepsilon} 
 \def\k{\kappa}      \def\l{\lambda}      
         \def\F{\Phi}        
 \def\j{\psi}            \def\r{\varrho}       \def\SS{\Sigma}
 \def\t{\tau}          
               
 \def\w{\omega}        

 \def\HH{{\mathcal H}}    \def\OO{{\mathcal O}}  
 \def\pa{\partial} \def\ra{\rightarrow}
  
 \def\dd{{\rm d}}  \def\bra{\langle}   \def\ket{\rangle}

 \def\iss{\ =\ }

 \def\fract#1#2{{\textstyle{#1\over#2}}}
 \def\ffract#1#2{\raise .2 em\hbox{$\scriptstyle#1$}\kern-.35em/
                 \kern-.2em\lower .15 em \hbox{$\scriptstyle#2$}}
 
 \def\half{\fract12}  
 
   \def\low#1{\raise-.3em\hbox{$\scriptstyle{#1}$}}
  
\def\bmatrix{\begin{matrix}} \def\ematrix{\end{matrix}} \def\bpmatrix{\begin{pmatrix}}\def\epmatrix{\end{pmatrix}}
\def\bcenter{\begin{center}} \def\ecenter{\end{center}}

\def\lowerheightfig#1#2#3{\(\raise-#1\hbox{\includegraphics[height=#2]{#3}}\)}
\def\lowerwidthfig#1#2#3{\(\raise-#1\hbox{\includegraphics[width=#2]{#3}}\)}
\def\qqquad{\qquad\qquad}

\def\glt{\hbox{\,\raise .35em\hbox{$>$}\kern-.8em\raise-.15em\hbox{$<$}\,}}

\publishversion
\begin{document} \begin{titlepage}

\title{\normalsize \hfill ITP-UU-13/29  \\ \hfill SPIN-13/21
\vskip 10mm \LARGE\bf Hamiltonian formalism for integer-valued variables and integer time steps\\[10pt]
\Large{and a possible application in quantum physics}}
\author{Gerard 't~Hooft}
\date{\normalsize Institute for Theoretical Physics \\
Utrecht University \\ and
\medskip \\ Spinoza Institute \\ Postbox 80.195 \\ 3508 TD Utrecht, the Netherlands \smallskip \\
e-mail: \tt g.thooft@uu.nl \\ internet: \tt
http://www.staff.science.uu.nl/\~{}hooft101/}

\maketitle
\begin{quotation} \noindent {\large\bf Abstract } \medskip \\
Most classical mechanical systems are based on dynamical variables whose values are real numbers.  Energy conservation is then guaranteed if the dynamical equations are phrased in terms of a Hamiltonian function, which then leads to differential equations in the time variable. If these real dynamical variables are instead replaced by integers, and also the time variable is restricted to integers, it appears to be hard to enforce energy conservation unless one can also derive a Hamiltonian formalism for that case. We here show how the Hamiltonian formalism works here, and how it may yield the usual Hamilton equations in the continuum limit. The question was motivated by the author's investigations of special quantum systems that allow for a deterministic interpretation. The `discrete Hamiltonian formalism' appears to shed new light on these approaches. 
\end{quotation}

\vfill \flushleft{\today}
\end{titlepage}

\newsecl{Introduction}{intro}
The existence of a conserved quantity called energy is a central concept in classical mechanics, closely related to Isaac Newton's principle of action = reaction: if the energy of one part of a system changes due to forces acting on it, an opposite change must occur somewhere else. An extremely elegant way to characterize forces where such an action principle is guaranteed is to relate these forces to the Hamiltonian\fn{Sometimes we see ``hamiltonian"  and ``lagrangian" written in lower case; we here decided to stick to the more natural looking ``Hamiltonian" with capital \textit{H}.} formalism\cite{hamilton}:
	\bi{-} Physical degrees of freedom are labeled as position variables \(q_i(t)\) and momentum variables \(p_i(t)\), where the index \(i\) can take any number \(n\) of values. The variables depend on one time variable \(t\). Thus, the dynamical variables are
		\be\{q_i,\,p_i\}\ ,\qquad i=1,\cdots,\,n\ . \eel{pqvariables} \ei
Typically, if we have \(N\) particles in a \(D\) dimensional space, \(n=ND\).  The time variable \(t\) will usually not be indicated explicitly.
	\bi{-} A Hamiltonian function \(H\) is defined in terms of the variables \(q_i\) and \(p_i\), equal to the total energy of the system:
		\be H(\vec q,\,\vec p\,)\ ,\eel{hamilton}
	where we wrote \(q_i\) and \(p_i\) as \(n\)-dimensional vectors.
	\itm{-} The evolution equations for the system are postulated to be
		\be {\dd q_i\over\dd t}=\dot q_i={\pa H(\vec q,\,\vec p\,)\over\pa p_i}\ ,\qqquad {\dd p_i\over \dd t}=\dot p_i=-{\pa H(\vec q,\,\vec p\,)\over\pa q_i}\ . \eel{hameqs}
	The dot here stands for differentiation with respect to time \(t\).
	\ei
In many cases, the Hamiltonian \(H\) can be written as the sum of a \(\vec p\)-dependent piece \(T\), a \(\vec q\)-dependent piece \(V\) and a term linear in \(\vec p\):
	\be H=T(\vec p\,)+V(\vec q\,)+p_iA_i(\vec q\,)\ , \qquad T(\vec p\,)=\sum_i{p_i^2\over 2m_i}\ ,\eel{HTV}
where the functions \(\vec A(\vec q\,)\) and \(V(\vec q\,)\) can be almost any differentiable function of \(\vec q\).

Since
	\be \dot H=\dot q_i{\pa H\over\pa q_i}+\dot p_i{\pa H\over\pa p_i}=-\dot q_i\dot p_i+\dot p_i\dot q_i=0\ , \eel{Hdot}
the energy \(E=H(\vec q,\,\vec p\,)\) is conserved in time.

A very important property of the Hamilton equations \eqn{hameqs} is that , by inverting the sign of \(\vec p\), one gets the equations backward in time, or, the system is time-invertible. This is a feature that we notice to exist in the real world. The planetary system, for instance, observes the same equations if we run these backwards in time. 

Now, we wish to consider a dynamical system in which all dynamical variables consist of integers, for instance pairs of integers \(Q_i\) and \(P_i\), and also the time variable \(t\) is always an integer,  so that the evolution in time must proceed over integer time steps \(\d t=1,\,2,\,\cdots\). One may be interested in an evolution law \(Q_i(t+1)=f_i(\,\vec Q(t),\,\vec P(t)\,)\) and \(P_i(t+1)=g_i(\,\vec Q(t),\,\vec P(t)\,)\), such that there exists some function \(H(\vec Q,\,\vec P)\) that is conserved in time. Also, we might desire time-invertibility for this discrete system as well. How should the functions \(f_i\) and \(g_i\) be chosen?

One might hope that replacement of the equations \eqn{hameqs} by difference equations might do the job, but of course that does not work, since the derivation \eqn{Hdot} would require higher order corrections that do not cancel out. Only if the functions \(f_i\) and \(g_i\) are strictly linear in \(\vec Q\) and \(\vec P\), one can find a conserved quadratic expression in \(\vec Q\) and \(\vec P\), but then, more often than not, this quadratic expression is not bounded below or above, and that cannot serve very well as an expression for energy. Indeed, the \emph{reason} why one often wishes to consider the conservation of energy is that energy consists only of positive contributions from the various parts of a system, so that if the total energy of a system is bounded, so are all its parts; this ensures stability of a system of equations.

There may be several reasons for being interested in discrete versions of the Hamiltonian formalism. One is that in numerical simulations often rounding errors may occur. These rounding errors might result in the energy not being exactly conserved. The total energy then usually increases, so that a system may run out of control. A discrete Hamiltonian formalism that ensures conservation of an energy that by itself is also an integer, can be handled numerically without rounding errors, so that conservation of energy would be guaranteed. This author, however, has another motivation: his search for theories underpinning quantum mechanics\cite{gthdeterm}.  Since that subject is controversial, and other researchers may have quite different reasons for being interested in the discrete Hamiltonian formalism, we postpone this motivation until the end of the paper, where this possible use of our findings and their implications are briefly discussed. In a separate paper, we plan to report on our work on deterministic interpretations of quantum mechanics with more detail.

\newsecl{One-dimensional system: a single $Q,\,P$ pair}{oned}
Our strategy will always be the same: we \emph{first} postulate an energy function \(H(\vec Q,\,\vec P)\) which must be integer valued  (usually, we shall indicate integers by capital letters).  And \emph{then} we search for the evolution law that keeps this quantity exactly conserved. 

In principle, it seems to be very simple to find such an evolution law: compute the total energy \(E\) of the initial state, \(H(\,\vec Q(0),\,\vec P(0)\,)=E\). Subsequently, search for all other values of \((\vec Q,\,\vec P)\) for which the total energy is the same number. Together, they form a subspace \(\SS_E\) of the \(\vec Q,\,\vec P\) lattice, which in general may look like a surface. Just consider the set of points in \(\SS_E\), make a mapping \((\vec Q,\,\vec P)\mapsto(\vec Q',\,\vec P')\) that is one-to-one, inside \(\SS_E\). This law will be time-invertible and it will conserve the energy. Just one problem then remains: how do we choose a unique one-to-one mapping?

The answer will be that we do the mapping sequentially: take the series of pairs \(Q_i,\,P_i\) for every value of the index \(i\) and update them for the index \(i\) taking the values \(1,\cdots, n\) one at the time. This reduces our problem to that of updating a single \(Q,\,P\) pair, such that the energy is conserved. This should be doable. Therefore, let us first consider a single \(Q,\,P\) pair.

There is a risk here: if the integer \(H\) tends to be too large, it will often happen that there are no other values of \(Q\) and \(P\) at all that have the same energy. Then, our system cannot evolve. So, we will find out that some choices of the function \(H\) are better than others. We shall see how this happens. 

Thus,  first consider a single pair, \(Q\) and \(P\). These variables form a two-dimensional lattice. Given the energy \(E\), the points on this lattice where the energy \(H(Q,P) =E\) form a subspace \(\SS_E\). We need to define a one-to-one mapping of \(\SS_E\) onto itself.

The case of a completely general integer-valued function \(H( Q, P)\) will still be too difficult for us. But the restricted case
	\be H( Q,P)=T( P)+V( Q)+ A(Q)\,B(P)\ , \eel{hamTVdiscr}
can be handled for fairly generic choices for the functions \(T(P ),\  V(Q),\ A(Q )\) and \(B(P )\). The last term here, the product \(A\,B\), is the lattice generalization of the magnetic term in Eq.~\eqn{HTV}. The function \(B(P )\) does not have to be linear in \(P\) if the kinetic term \(T(P )\) deviates from being purely quadratic, being restricted to integer values, while (as will be seen later) the function \(T_0(P )=P^2\) itself will often be too steep. Many interesting physical systems, such as most many body systems, will  be covered by Eq.~\eqn{hamTVdiscr}. Often, we shall disregard the last term.

	\begin{figure}[h!] \setcounter{figure}{0} \begin{quotation}\begin{center}
			\lowerwidthfig{0pt}{140mm}{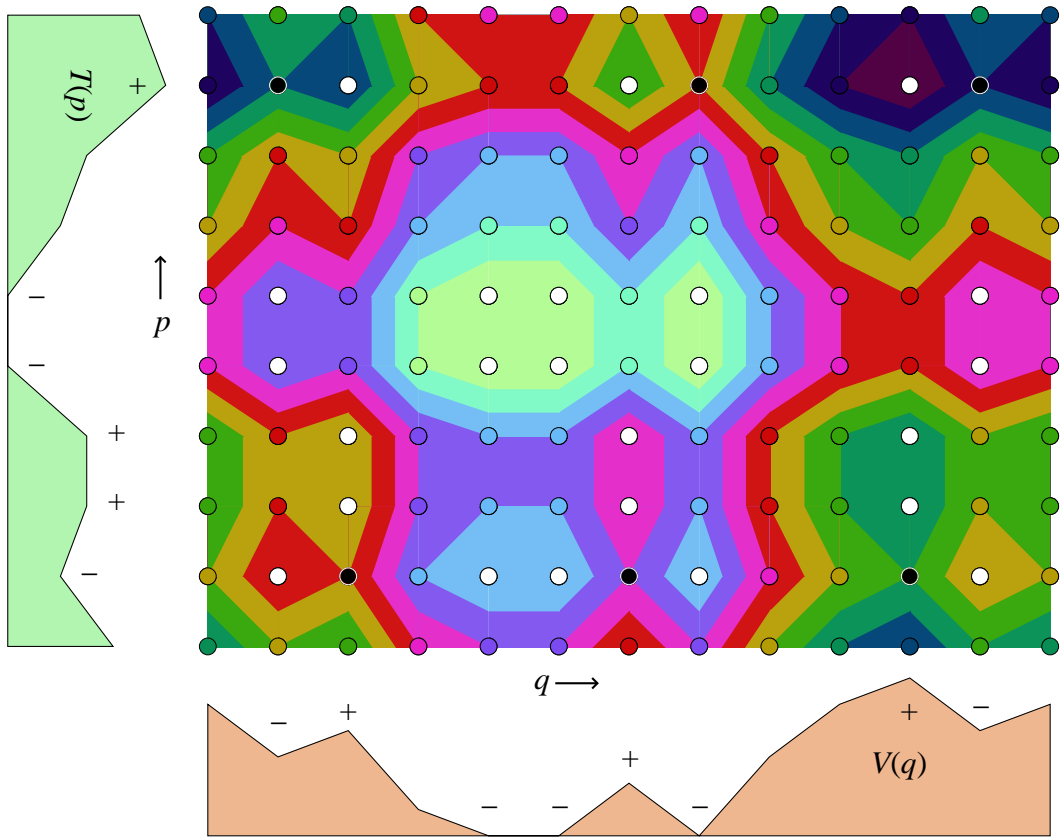}
 \caption{\small The \(Q\,P\) lattice in the 1+1 dimensional case. Constant energy contours are here the boundaries of the differently colored regions. Points of the lattice on the contours are indicated in the same colors. White points are not on a contour and therefore these are stable rest points.  Black spots are on saddle points, having two contours there. These will also be declared to be stationary points. To establish how the contours themselves go through these points, we lift the energy by some infinitesimal amount (see text). The exceptional points are related to local maxima  \((+)\) and minima \((-)\) of the functions \(T\) and \(V\).\labell{PQcontours.fig}}
\end{center} \end{quotation}
		\end{figure}

Finding the invertible mapping is now done as follows. First, we extrapolate the functions \(T,\ P,\ A\) and \(B\) to all real values of their variables. Write  real numbers \(q\) and \(p\) as
	\be q=Q+\a\ , \quad p=P+\b\ ,\qquad Q\hbox{ and } P\hbox{ integer,}\quad 0\le\a\le 1\ ,\quad 0\le\b\le 1\ . \eel{realintfract}

Then define the continuous functions 
	\be V(q)=(1-\a)V(Q)+\a V(Q+1)\ ,\qquad T(p )=(1-\b)T(P )+\b T(P+1)\ , \eel{VTcont}
and similarly \(A(q)\) and \(B(p )\). Now, the spaces \(\SS_E\) are given by the lines  \(H(q,p)=T(p )+V(q)+A( q)B(p )=E\), which are now sets of oriented, closed contours, see Fig.~\ref{PQcontours.fig}. They are of course the same closed contours as in the standard, continuum Hamiltonian formalism.

The standard Hamiltonian formalism would now dictate how fast our system runs along one of these contours. We cannot quite follow that prescription here, because at \(t\,=\) integer we wish \(P\) and \(Q\) to take integer values, that is, they have to be at one of the lattice sites. But the speed of the evolution does not affect the fact that energy is conserved. Therefore we modify the speed, by now postulating that
\begin{quote} \emph {at every time step \(t\ra t+1\), the system moves to the next lattice site that is on its contour \(\SS_E\).} \end{quote}
If there is only one point on the contour, which would be the state at time \(t\), then nothing moves. If there are two points, the system flip-flops, and the orientation of the contour is immaterial. If there are more than two points, the system is postulated to move in the same direction along the contour as in the standard Hamiltonian formalism. In Fig.~\ref{PQcontours.fig}, we see examples of contours with just one point, and contours with two or more points on them. Only if there is more than one point, the evolution will be non-trivial.

In some cases, there will be some ambiguity. Precisely at the lattice sites, our curves will be non-differentiable because the functions \(T,\ C,\ A,\) and \(B\) are non-differentiable there. To lift the ambiguity, we replace the contours \(\SS_E\) by \(\SS_{E+\e}\), where \(\e>0\) is infinitesimally small. This will make the contours unique in most cases, but it may happen that, near a lattice site, now \emph{two} contours emerge. In that case, these will go in opposite directions, and so they cancel one another out. This cancellation will leave us with acceptable contours only, see Fig.~\ref{PQcontours.fig}. At the extremal points of the Hamiltonian (white points in the Figure), as well as in the saddle points (black points), the evolution is declared to stand still, but the continuation of the contours is still prescribed by the \(\e\) prescription. All this is necessary to make the evolution law time-invertible. The fact that there are stationary points is not problematic if this description is applied to formulate the law for multi-dimensional systems, see Section~\ref{highdim}.

At first sight, it might seem that our discrete Hamiltonian prescription departs quite a bit from Hamilton's equations for continuous systems, but the departure is not greater than what one might expect from discretization. Observe that the odds that a curve in Fig.~\ref{PQcontours.fig} hits a lattice site are inversely proportional to the gradients of the functions \(T\) and \(P\). Therefore, the average distance between two adjacent lattice points on a contour is approximately equal to that gradient. This means that the system will move across the lattice at speeds that, on average, stay close to the gradient of the Hamiltonian.

Indeed this situation resembles the one in the standard Hamilton equations, where the speed at which a state point moves in phase space is also given by the absolute value of the gradient of the Hamiltonian function in that point in phase space. As an exercise, consider the case \def\intt{\mathrm{int}}
	\be T(P )=\intt(P^\k/2m)\ ,\quad V(Q)=\intt(\l Q^\g)\ ,\quad A=B=0\ ,\quad 0<\k,\g<2\ , \eel{TVABexample}
where `int' stands for the integer part, or `floor' of the subsequent expression.
By counting lattice sites, and taking the number of contours within a given contour to be equal to \(E\), one estimates the average number of lattice sites on a given contour to be close to
	\be n(E)\approx \l^{-1/\g}(2m)^{1/\k}E^{\,1/\k\,+\,1/\g\,-1}\ , \eel{contourpoints} 
while the continuum Hamiltonian formalism would tell us that the amount of time needed to traverse the contour would be roughly
	\be \t(E)\approx {|p|\over|{\pa H\over\pa q}|}\Bigg|_{\mathrm{average}}\approx\quad{|q|\over|{\pa H\over\pa p}|}\Bigg|_{\mathrm{average}}					\approx\ C(\k,\g)\,n(E)\ , \eel{match}
where \(C=\OO(1)\). In our estimate, it is not exactly one, but we expect the average value to approach to a universal constant exactly. We note that our estimated expressions only make sense if \(1/\k+1/\g>1\), so that either \(\k\) or \(\g\) or both must be smaller than 2.
Our discrete Hamilton procedure works best if \(n(E)\) in Eq.~\eqn{contourpoints} tends to stay greater than one, otherwise the system comes to a standstill. 

This example also tells us that there is a practical restriction on the functions \(T\), \(V\), \(A\) and \(B\) in Eq.~\eqn{hamTVdiscr}: these functions must be sufficiently smooth; if they vary too wildly, all contours will have only one point on them so that nothing moves. We cannot turn this condition into a rigorous inequality, since no harm is done if a system comes to rest occasionally (in a multi-dimensional system it will be kicked out of such points soon enough). A discrete integer function such as \(V(Q)\) is `smooth' if its discretised derivative in \(Q\) does not switch sign too frequently. In many cases, a condition such as
	\be |V(Q_1)-V(Q_2)|<|Q_1-Q_2|\,(|Q_1|+|Q_2|)\ , \eel{workable}
and similar constraints on \(T,\ A\), and \(B\) will suffice.

This completes the 1+1 dimensional case. We found an evolution law that exactly preserves the discrete energy function chosen. The procedure is unique as soon as the energy function can be extended naturally to a continuous function between the lattice sites, as was realized in the case \(H=T+V+AB\) in Eq.~\eqn{VTcont}. Furthermore we must require that the energy function does not vary too steeply, so that most of the closed contours contain more than one lattice point. This excludes most purely quadratic functions of the integers \(P\) and \(Q\), since then, in Eq.~\eqn{contourpoints}, \(\k=2,\ \g=2\) while
\(\l\) and \(1/2m\) must be integers.

\newsecl{The multi dimensional case}{highdim}

The 1+1 dimensional case, as described in the previous section, is rather boring, since the motion occurs on contours that all have rather short periods.  In higher dimensions, this will be very different. So now, we consider the variables \(Q_i,\ P_i,\ i=1,\cdots,n\). Again, we postulate a Hamiltonian \(H(\vec Q,\vec P\,)\) that, when \(P_i\) and \(Q_i\) are integer, takes integer values only. Again, let us take the case that
\be H(\vec Q,\,\vec P\,)=T(\vec P\,)+V(\vec Q\,)+A(\vec Q\,)\,B(\vec P\,)\ . \eel{HTVABvec}
To describe an energy conserving evolution law, we simply can apply the procedure described in the previous section \(n\) times for each time step. For a unique description however, it is now mandatory that we introduce a \emph{cyclic ordering} for the values \(1,\cdots,n\) that the index \(i\) can take. Naturally, we adopt the notation of the values for the index \(i\) to whatever ordering might have been chosen: 
	\be 1<2<\,\cdots\,<n<1 \,\cdots\ . \eel{cyclic}
We do emphasize that the procedure described next depends on this ordering.

Let us denote the operation in one dimension, acting on the variables \(Q_i,P_i\) at one given value for the index \(i\), as \(U_i\). Thus, \(U_i\) maps \((P_i,Q_i)\mapsto(P'_i,Q'_i)\) using the procedure of Section~\ref{oned} with the Hamiltonian  \eqn{HTVABvec}, simply keeping all other variables \(Q_j,P_j,\ j\ne i\) fixed. By construction, \(U_i\) has an inverse \(U_i^{-1}\). Now, it is simple to produce a prescription for the evolution \(U\) for the entire system, for a single time step \(\d t=1\):
	\be U(\d t)=U_n\,U_{n-1}\,\cdots\,U_1\ . \eel{totalU}
where we intend to use the physical notation:   \(U_1\) acts first, then \(U_2\), etc., although the opposite order can also be taken. Note, that we have some parity violation: the operators \(U_i\) and \(U_j\) will not commute if \(i\ne j\), and therefore the resulting operator \(U\) is not quite the same as the one obtained when the order is reversed. 

Time inversion gives:
	\be U(-\d t)=U(\d t)^{-1}=U_1^{-1}U_2^{-1}\cdots U(n)^{-1}\ . \eel{timereversal}
The product \(P\)(parity)\,\(T\)(time inversion) may still be a good symmetry. We believe that, in the real world, this corresponds to CPT symmetry, while \(P\), \(T\), or \(CP\) are not respected. 

\newsecl{Cellular automata}{CA}
	A cellular automaton\cite{wolfram} is a system consisting of a \(D\)-dimensional array of memory cells, each containing a small amount of data. At the beat of a clock, the data in each cell are updated according to a fixed prescription, depending on the contents of its own data and the data in neighboring cells only. As time goes on, information usually expands with a given speed through the entire system in all directions.
	
	One may require that the evolution law of an automaton be time-invertible. This can be achieved by the so-called Margolus rule\cite{margolus}, where each cell remembers its own data and the data it had one time step earlier. The evolution rule followed by each cell is that it follows an invertible law such as addition or multiplication -- let us say it is addition -- to obtain its contents at time \(t+1\) by adding a given function of the neighboring cells and itself at time \(t\) to its own data at time \(t-1\). Such a rule is sufficiently generic to be applicable as a model for many \(D\)-dimensional dynamical systems, and it is easy to reverse in time.
	
	However, the Margolus prescription does not easily allow for the construction of a conserved energy concept in the form of a non negative integer number. This implies that, in practice, time-invertible Margolus cellular automata are very unstable; they quickly produce completely pseudo-random configurations. 
	
	Now, we propose a different class of cellular automata, which is the class of automata that obeys a discrete Hamiltonian principle. We take the quantity that serves the role of the index \(i\) in the previous Section, to take the form of a discrete space-like coordinate \(\vec x\), with in addition possibly more than one internal components: \(i\ra(\vec x,\,k)\), with the ordering, as discussed in the previous Section, to be specified shortly. The integer-valued variables \(Q_i,\,P_i\) of the previous Section are now replaced by integer valued fields \(\F_k(\vec x)\) and \(P_k(\vec x)\), the latter being the canonical momenta of the fields \(\F_k(\vec x)\). Our Hamiltonian  will now be the Hamiltonian of this field system. In principle, this Hamiltonian is constructed and treated the same way as in more conventional field theories in the continuum. We could write for example \\		\vbox{
	\be && H\iss\sum_{\vec x}\HH(\vec x)\ ,\qquad \HH(\vec x)\ = \cr
	&&\intt\bigg(\half\r \sum_{k}\bigg(\sum_{a=1}^D((\F_k(\vec x+\vec e_a)-\F_k(\vec x))^2+ m_k^2\F_k^2(\vec x)\bigg)\bigg)
	 +\intt\bigg(\half\r\sum_kP_k^2(\vec x))\bigg)\ ,\qquad\hbox{ } \eel{discfieldham}}
where  \(\vec e_a\) are the unit vectors in the direction \(a\), and the factor \(\r\le 1/D\) is necessary in order to reduce the coefficient for the quadratic terms sufficiently so as to get Hamiltonian contours with a sufficient number of points on them. It does imply that the quantized time unit is smaller than the quantized space unit.

We now specify the (cyclic) order at which the numerous variables of the system (the fields \(\F_k(\vec x)\) and their momenta \(P_k(\vec x)\) at all positions \(\vec x\) and all values of \(k\)) have to be updated: \emph{first} update all the even sites \(\vec x\) and \emph{then} all odd sites. Our construction, writing the Hamiltonian as a sum over Hamilton densities \eqn{discfieldham}  \emph{after} taking the integer parts ensures that the updating operators \(U_{\vec x}\) commute with the updating operators \(U_{\vec x\,'}\) as soon as \(\vec x\) and \(\vec x\,'\) are further apart than nearest neighbors, for instance when they are both on the sub lattice of the even sites or both on the sub lattice of the odd sites. So, while we update all even sites, the result only depends on nearest neighbors. This ensures that information does not travel faster than the speed of light --- if we define the speed of light to be equal to the size of the lattice length in space divided by the lattice length in time.

Note that, on the one hand, the model described in Eq.~\eqn{discfieldham} is, in a sense, the discrete version of a field theory where the speed of light would be \(D\) times lower, since the Hamiltonian was multiplied by \(\r\le 1/D\). On the other hand, the discretization procedure applied here does not respect Lorentz invariance, so there is no contradiction. Our model was not intended to be a credible model of the real world but just an example of a field theoretical system that can be handled as a toy model. It is an improvement in comparison with the cellular automata we described in earlier publications because models of this sort have an absolutely conserved energy, and are based on a Hamiltonian procedure, just as the continuous models normally considered in physics. More experience in the study of the new models will be needed.

One may well ask why we should want to pay a rather big price for having a conserved energy. The price is big because the explicit calculation of the contours and the identification of the next lattice points on them may be  cumbersome numerically. We think that the discrete Hamiltonian formalism will be important. One very suggestive application will be the role it may play in the discussion of quantum systems with underlying determinism. This we briefly explain in the next Section.

\newsecl{Discrete Hamiltonian formalism in a theory of local hidden variables}{localhidden}

It is generally believed that theories of local hidden variables are incompatible with known features of quantum mechanics\cite{nohiddenv}, unless one manages to employ the loophole called ``super determinism"\cite{superdet}, which is the notion that everything that happens in the universe is completely determined by the evolution laws, and that for instance ``counterfactual" experiments\cite{counterfact} or measurements are impossible. Even if one can use that loophole, one still hits upon very strange features. For instance when an experimenter observes a photon through a polarization filter, the choice (s)he makes for the angle of the filter was already part of the ontological description of the photon when it was produced, which could have been many years ago in a distant star. Many researchers consider this as an unacceptable feature and without much further ado dismiss such theories. Bell's inequalities\cite{bellineq} do just more of the same, turning the apparent contradiction in a contradiction of numbers. 

It may therefore come as a surprise that nevertheless models exist that combine quantum mechanics with classical behavior. We have searched for a model with the following properties:
	\bi{$i$} The model is described by classical equations that locally determine how the system evolves in time.
	\itm{$ii$} At the same time the physical states span a Hilbert space, in which the evolution law takes the form of Schr\"odinger's equation,
		\be {\dd\over\dd t}|\j(t)\ket=-iH|\j(t)\ket\ , \eel{schro}
	\itm{$iii$} The Hamiltonian \(H\) can be written as the integral (or sum) over space of a Hamilton density \(\HH(\vec x)\), obeying local commutation rules: there is an \(\e>0\) such that at fixed time \(t\) we have
		\be [\HH(\vec x),\,\HH(\vec x')]=0\ ,\quad \hbox{if } \  |\vec x- \vec x'|>\e\ . \eel{localhamdensity}
	\itm{$iv$} The Hamiltonian has a lower bound, which we can put at zero: \(\bra\j|H|\j\ket\ge 0\).
	\itm{$v$} The Hamiltonian does not have an upper bound.
	\ei
This last condition will be important if gravity is coupled to our system. An upper bound of the Hamiltonian would naturally occur in the Planck domain, and although such an upper bound would be large when compared to typical quantum systems studied under earth-like conditions, it would be far too low to understand the gravitational force, which acts on energy/mass  density. Curvature of space and time due to sources in the domain of the Planck mass, being a few tens of micrograms, will be insignificant in most cases. 

We now have a model that nearly, but not quite, obeys these requirement. Strangely enough, condition $iii$ has to be modified. It is not fulfilled but we \emph{do} have locality! Our models are of the class described in the previous Section, Eq.~\eqn{discfieldham}. \def\fr{{\mathrm{fract}}}

The field variables \(\F_k(\vec x)\) and \(P_k(\vec x)\) are discrete integers spread over a discrete lattice of coordinates \(\vec x\). We may consider Hilbert space spanned by the states in which all these fields have specified values. In this Hilbert space, the evolution law over one fundamental time step \(\d t\), which for simplicity we normalize to 1 rather than \(1/D\),  can be regarded as a unitary operator \(U=U(\d t)\). It has eigen values \(e^{-i H_\fr}\) where the real number \(H_\fr\) lies in the interval \(-\pi<H_\fr<\pi\). It is conserved in time.

Also conserved in time is the Hamiltonian in Eq.~\eqn{discfieldham}. Since this Hamiltonian is built from integers, we call it \(H_\intt\). The total Hamiltonian is now a real number operator, uniquely defined by
	\be H=2\pi H_\intt+H_\fr+\pi\  . \eel{Htot}
This defines our model.

The operator \(H_\fr\) can be calculated from \(U(\d t)\) as follows. By Fourier transformations, one easily derives that, if \(-\pi<\w<\pi\),
	\be \w=2\sum_{n=1}^\infty{(-1)^{n-1}\sin(n\w)\over n}\ , \eel{omegasumsin}
so that
	\be H_\fr=\sum_{n-1}^\infty{(-1)^{n-1}\over n\,i}(U(n\,\d t)-U(-n\,\d t))\ . \eel{HfrUn}

This sum converges nearly everywhere, but it is not quite local, since the evolution operator over \(n\) steps in time, also acts over \(n\) steps in space. Both \(H_\intt\) and \(H_\fr\) are uniquely defined, and since \(H_\fr\) is bound to an interval while \(H_\intt\) is bounded from below, also \(H\) is bounded from below.

We do note that demanding a large number of low energy states near the vacuum (the absence of a large mass gap) implies that \(U(n\,\d t)\) be non-trivial in the \(H_\intt=0\) sector. This is not the case in the model  \eqn{discfieldham}; it would probably require a more complicated space of field variables, but in principle there is no reason why such models should not exist also. In fact, the cellular automaton models discussed in Ref.~\cite{CAGtH} have no manifest conserved \(H_\intt\), so that all their states can be regarded as sitting in the \(H_\intt=0\) sector of the theory.

Because of the non-locality of Eq.~\eqn{HfrUn}, the Hamiltonian does not obey the rule \(iii\), but since \(U(\d t)\) is local, the evolution over integer time steps \(n\,\d t\) \emph{is} local, so the theory can be claimed to obey locality; we simply haven't defined its states at time \(t\) when \(t\) is not an integer.

If we could claim that in the physically relevant sector of Hilbert space the sum~\eqn{HfrUn} converges rapidly, we could argue that the \(\e\) defined in condition \(iii\) can be kept small. However, the sum does not converge rapidly everywhere in Hilbert space. We are interested in the Hamiltonian as it acts on states very close to the vacuum, in our notation: \(H_\intt=0,\quad H_\fr=\a-\pi\), where \(0<\a\ll\pi\). Suppose then that we introduce a cut-off in the sum \eqn{omegasumsin} and \eqn{HfrUn} by multiplying the summand with \(e^{-n/R}\), where \(R\) is also the range of non-locality of the last significant terms of the sum.

In Eq.~\eqn{omegasumsin}, the variable \(w=\a-\pi\). The sum with cut-off can be evaluated exactly:
	\be \a-\pi\ \approx\ 2\sum_{n=1}^\infty{(-1)^{n-1}\sin(n\,\w)e^{-n/R}\over n} &=&
		2\arctan \bigg({\sin\w\over e^{1/R}+cos\w}\bigg) \cr 
		\iss-\pi+2\arctan\bigg({\cos\w+1+\ffract{1\,}{R}\over\sin\w}\bigg)	&=&
		-\pi+2\arctan\bigg({\sin\a\over 1+\cos\a}+{\ffract{1\,}R\over\sin\a}\bigg)\ ,\quad\hbox{\ }		\eel{truncsum}
where we replaced \(e^{1/R}\) by \(1+\ffract{1\,}R\) since \(R\) is large and arbitrary. 

Writing \({\sin\a\over 1+\cos\a}=\tan\half\a\), we see that the approximation becomes exact in the limit \(R\ra\infty\). 
We are interested in the states close to the vacuum, having  a small but positive energy \(H=\a\).
Then, at finite \(R\), the cut-off at \(R\) replaces the Hamiltonian \(H_\fr\)  by 
	\be H_\fr\ra H_\fr+{2\over R\,H_\fr}\ , \eel{Happrox}
and this is only acceptable if
	\be R\gg M_{\mathrm{Pl}}/\bra H_\fr\ket ^2\ . \eel{convradius}	
Here, \(M_{\mathrm{Pl}}\) is the ``Planck mass", or whatever the inverse is of the elementary time scale in the model. This cut-off radius \(R\) must therefore be chosen to be  very large, so that, indeed, the exact quantum description of our local model generates non-locality in the Hamiltonian. One might speculate that it is this non-locality that could explain away the apparent disagreements with what was thought of as ``common sense".

A big step forward compared to our earlier discussions of cellular automata as models of quantum physics is that now our expressions contain only convergent sums; previously, we had to truncate divergent expressions while we did not know how convergent or divergent they were. Previously, we had assumed that the ``Planck mass"  \(M_{\mathrm{Pl}}\) is so large that, in quantum theories, no domains of higher values for \(H_\intt\) than the trivial one are needed. Now that we know, in principle, how to handle the less trivial Sectors \(H_\intt=1,2,\cdots \infty\).                                                                                             

As this section is just a summary of our ideas about quantum determinism, we do not pretend that the discussion of this section would be complete; more will appear elsewhere, in due time.	Let us emphasize one thing clearly, since `super determinism' raises much suspicion in general: there is no spooky acausality, or `retro-causality', of any sort in the classical description of our models.

\end{document}